# Interpretable Activation-Selection Neural Networks for Symbolic Regression of Parameter-Dependent Hamiltonian Eigenvalues

Alexander Yu. Kharin[1]*, Kirill F. Sheberstov[2]*

[1] *InfoMediji LTD*

[2] *Chimie Physique et Chimie du Vivant (CPCV, UMR 8228), Department of Chemistry, École Normale Supérieure, PSL University, Sorbonne University, CNRS, Paris, 75005, France*

Contact authors:

alexankhar@infomediji.com

kirill.sheberstov@ens.psl.eu

ORCID:

AYK: 0000-0002-0147-0384

KFS: 0000-0002-3520-6258

# Abstract

Analytical approximations to eigenvalues of parameter-dependent Hamiltonians can provide physical insight that is not readily apparent from numerical diagonalization alone. Here, we introduce an activation-selection network (ASN), a differentiable symbolic-regression architecture in which each input node learns a sparse combination of predefined analytical functions and the trained network can be converted directly into an explicit expression. Before regression, the Hamiltonian parameters and eigenvalues are expressed as dimensionless ratios. This normalization enforces dimensional homogeneity, reduces the number of independent variables, and ensures that the extracted expressions do not depend on the choice of energy units. Using the library $\{0, x, x^2\}$, compositions across successive hidden layers generate polynomial expansions of progressively higher degree; polynomial expansions of arbitrary finite degree can therefore be obtained in principle by increasing the network depth. We apply the ASN to effective three- and four-site spin-chain Hamiltonians relevant to zero-quantum nuclear magnetic resonance. Comparisons with degenerate perturbation theory show that the extracted expressions capture the expected constant, linear, and quadratic structure. Fixed-basis least-squares models match or slightly outperform the ASN when an appropriate quadratic basis is specified in advance, while inclusion of a radial feature improves the local approximation near the degeneracy. These results establish the ASN as a differentiable framework for selecting compact symbolic representations when several functional forms are plausible, while showing that adaptive activation selection does not provide an intrinsic accuracy advantage over a suitable predefined basis.

# I. Introduction

Feedforward neural networks are universal approximators of continuous functions [1] and are widely used to model physical systems [2]. Their predictive flexibility, however, does not by itself provide compact analytical expressions relating physical variables. Symbolic regression pursues a different objective by seeking explicit mathematical expressions that balance approximation accuracy and structural simplicity. In physics, such expressions can reveal scaling laws, symmetries, limiting behavior, and conservation laws that may remain obscured in purely numerical models. Earlier differentiable approaches to equation learning include Equation Learner (EQL) models, which combine predefined analytical operators with sparsity regularization to identify interpretable equations [3,4]. More recently, Kolmogorov-Arnold networks (KANs) introduced a different neural architecture based on learnable univariate functions [5].

Specifically, KANs replace the scalar weights of conventional multilayer perceptrons (MLPs) with trainable univariate functions placed on the connections between nodes. In the original KAN framework, these edge functions are represented by splines and can subsequently be matched to and replaced by functions from a predefined symbolic library [5]. The activation-selection network (ASN) introduced here follows a different construction. For each node serving as an input to a layer, training selects one function from a predefined set of analytical functions. The resulting transformation is shared among all outgoing connections from that node, while their scalar weights are trained independently. The trained ASN therefore has an explicit analytical form and does not require a separate symbolic approximation of learned spline functions. This construction is suited to physical problems for which numerical solutions can be calculated readily, whereas compact analytical approximations remain difficult to derive.

The effectiveness of an ASN therefore depends on selecting a function library consistent with the expected analytical structure of the target problem. For Hamiltonian eigenvalue problems, perturbation theory provides such guidance when the Hamiltonian can be separated into a dominant part and one or more smaller perturbations [6]. The eigenvalues can then be expanded in powers of the corresponding perturbation parameters, with truncation of these series yielding polynomial approximations. This motivates the polynomial function library used in the present work. The ASN framework is not, however, intrinsically restricted to polynomial functions. Exact diagonalization near level crossings or avoided crossings can produce square-root dependencies, while ordering the eigenvalues can introduce absolute-value dependencies, even when the underlying Hamiltonian is analytic in its parameters.

Here, we apply the ASN to symbolic regression of parameter-dependent Hamiltonian eigenvalues. The network uses the activation set $\{0, x, x^2\}$, where 0 denotes the zero function, and compositions across successive layers can generate polynomials of increasing degree. As controlled benchmarks, we consider effective spin-chain Hamiltonians relevant to zero-quantum nuclear magnetic resonance (NMR) [7–9]. Their eigenvalues can be calculated numerically to provide exact reference data, whereas compact parameter-dependent analytical approximations become increasingly difficult to derive for longer or nonuniform chains. Expressing the Hamiltonians in terms of dimensionless ratios of original parameters reduces the number of independent variables and ensures dimensional homogeneity. For the uniform three-site chain, we compare the ASN expressions with results from degenerate perturbation theory. We then derive a two-parameter perturbative reference for the nonuniform three-site chain and extend the ASN analysis to a four-site chain. Fixed-basis least-squares polynomial fits provide an additional baseline for these comparisons. Together, these benchmarks test whether learned activation selection can yield compact analytical approximations beyond direct perturbative and least-squares approaches.

# II. Activation-selection neural-network framework

The goal of symbolic regression is to represent a target relationship using a mathematical expression optimized jointly for predictive accuracy and simplicity. The ASN implements this objective by combining a differentiable choice among predefined analytical activation functions with linear transformations between successive layers (**Fig. 1**).

The structure of a single ASN layer is illustrated in **Fig. 1(a)**. Each layer comprises an activation-selection sublayer followed by a linear-transformation sublayer. The activation-selection sublayer maps each input $x_i$ to a transformed variable $v_i$ by forming a differentiable combination of $k$ predefined activation functions. The subsequent linear-transformation sublayer combines these variables to produce the outputs $y_l$. The activation-selection parameters and linear weights are optimized jointly.

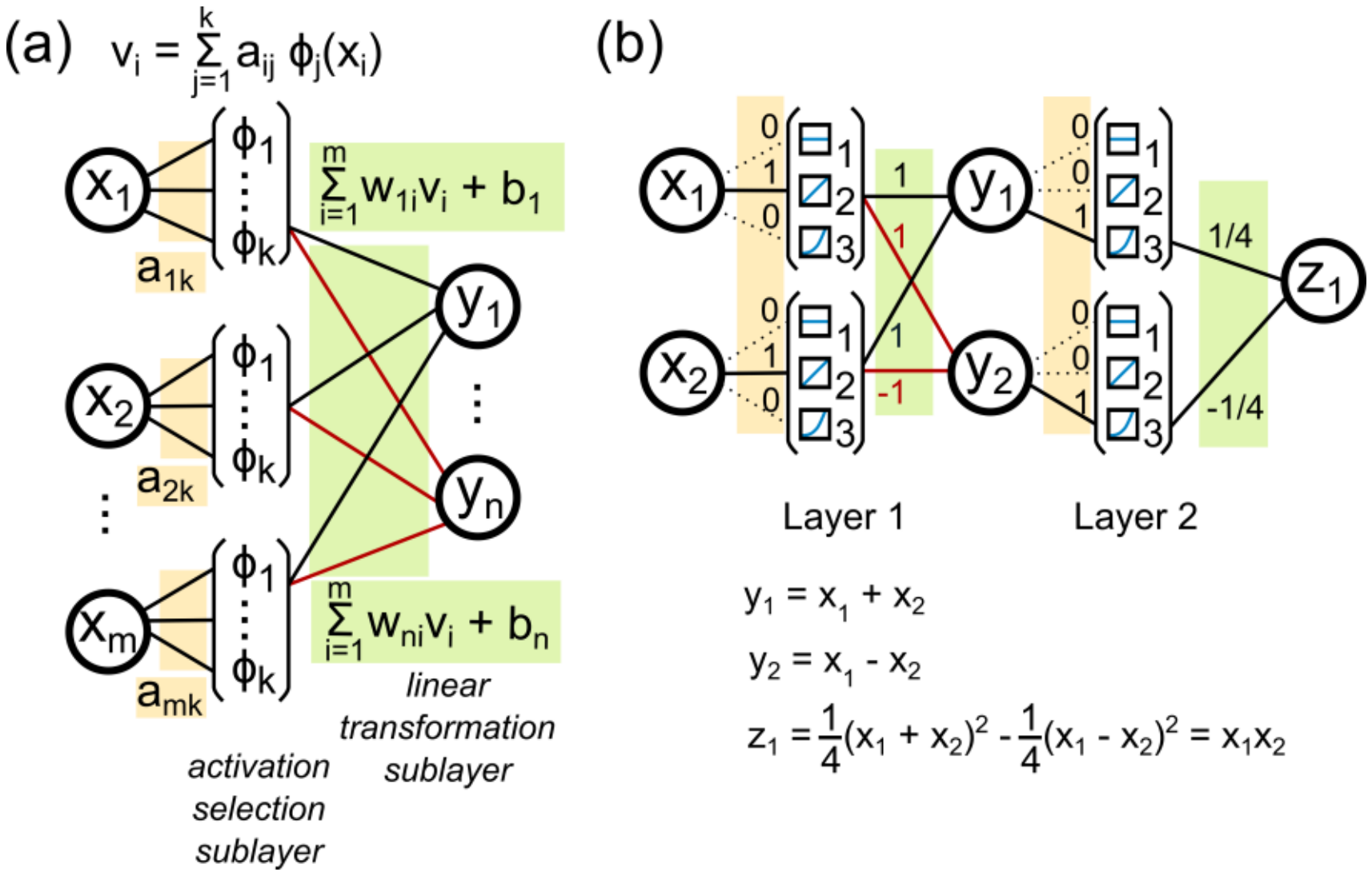

**Fig. 1. Architecture of the activation-selection network (ASN)**. (a) Structure of a single ASN layer. For each input $x_i$, $k$ candidate activation functions $\phi_j(x_i)$ are evaluated and combined using the softmax-normalized coefficients $a_{ij}$. An $L_{0.5}$-type penalty favours nearly one-hot coefficient vectors, thereby promoting the selection of a single activation for each input node. The transformed variables $v_i$ are then combined by a linear transformation with weights $w_{li}$ and biases $b_l$ to produce the output nodes $y_l$. (b) Two-layer ASN that exactly implements multiplication of two inputs. The first layer forms $y_1 = x_1 + x_2$ and $y_2 = x_1 - x_2$. The second layer selects the quadratic activation and combines the resulting terms as $z_1 = \frac{1}{4}(x_1 + x_2)^2 - \frac{1}{4}(x_1 - x_2)^2 = x_1x_2$. The candidate set $\{0, x, x^2\}$, where 0 denotes the zero function, permits higher-degree polynomials to be constructed by stacking layers and allows selected inputs to be suppressed.

To define the activation-selection step mathematically, each input node $x_i$ is assigned a vector of trainable logits $\boldsymbol{\alpha}_i = (\alpha_{i1}, \dots, \alpha_{ik})$, with one logit for each candidate activation function. Training proceeds in two stages. During the initial stage, the logits are converted into activation-selection coefficients using the softmax transformation,

$$a_{ij} = \mathrm{softmax}(\boldsymbol{\alpha}_i)_j = \exp(\alpha_{ij}) / \left[\sum_{r=1}^{k} \exp(\alpha_{ir})\right]. \tag{1}$$

The activation-selection coefficients are nonnegative and normalized in both stages:

$$a_{ij} \geq 0, \sum_{j=1}^{k} a_{ij} = 1. \tag{2}$$

The softmax parameterization allows all candidate activation functions to contribute through differentiable coefficients, enabling the logits to be optimized jointly with the linear weights. To favor nearly one-hot coefficient vectors, an $L_{0.5}$-type regularization term is minimized for each input node during the softmax stage:

$$R_{0.5}^{(i)} = \sum_{j=1}^{k} \sqrt{a_{ij}} - 1. \quad (3)$$

The total regularization term is obtained by summing $R_{0.5}^{(i)}$ over all input nodes in all ASN layers. Because the coefficients are nonnegative and sum to unity, the sum of their square roots approaches its minimum when one coefficient approaches unity and the remaining coefficients approach zero. Subtracting 1 sets this limiting minimum to zero without changing its location or the gradients of the regularization term.

During the final stage of training, softmax is replaced by sparsemax [10]:

$$a_{ij} = \mathrm{sparsemax}(\boldsymbol{\alpha}_i)_j = \max(\alpha_{ij} - \tau(\alpha_i), 0). \quad (4)$$

Here, $\tau(\alpha_i)$ is a threshold calculated independently for each input $i$ such that the coefficients satisfy Eq. (2). Unlike softmax, sparsemax can assign coefficients of exactly zero to unused activation functions, thereby producing a sparse analytical representation.

The $L_{0.5}$-type regularization favors coefficient vectors close to a vertex of the simplex, where one coefficient equals 1 and all others equal 0. Sparsemax can assign exact zeros to unused activation functions, but it guarantees sparsity rather than selection of a single function: several coefficients may remain nonzero. A final support size of one therefore corresponds to the selection of a single library function, such as 0, $x$, or $x^2$, whereas a larger support indicates a sparse mixture of several functions. We distinguish these two cases and report the final support sizes for the independent training runs discussed below.

The effective activation corresponding to input $x_i$ is given by:

$$v_i = \sum_{j=1}^{k} a_{ij} \phi_j(x_i). \quad (5)$$

Finally, the linear-transformation sublayer combines these activations using trainable weights and biases, as in a conventional feedforward layer. For an ASN layer with $n$ output nodes, the output $y_\ell$ is given by:

$$y_\ell = \sum_{i=1}^{m} w_{\ell i} v_i + b_\ell, \ \ell \in \{1, \ldots, n\}. \quad (6)$$

**Fig. 1(b)** demonstrates that the ASN can represent the product of two arbitrary input variables exactly using only the activation set {0, $x$, $x^2$}. Because the output of one ASN layer serves as the input to the next, repeated application of the available transformations can generate products and progressively higher powers. For an ASN with L layers, the polynomial degree can at most double at each layer, giving an upper bound of $2^L$. A two-layer ASN can therefore represent quartic terms in principle. In the specific construction shown in **Fig. 1(b)**, however, the first layer uses identity activations to form linear combinations, and only the second layer applies the quadratic activation. Explicit logits, weights, and biases for this construction are given in Appendix A. The construction in **Fig. 1(b)** is not unique: wider networks can represent the same product, while unnecessary paths can be suppressed by selecting the zero activation.

# III. Spin-chain Hamiltonians and benchmark eigenvalue problems

## A. Effective-spin representation and Hamiltonian

To evaluate whether the ASN can recover useful analytical approximations to Hamiltonian eigenvalues directly from numerical data, we consider effective spin-chain Hamiltonians describing coupled methylene proton pairs in zero-quantum NMR experiments [7–9]. In these systems, a local imbalance between the $T_0$ and $S_0$ populations can propagate along the molecular chain, providing an experimental realization of effective spin-chain dynamics [9]. The relevant eigenvalue differences determine the frequencies of the observed zero-quantum transitions, directly linking the regression targets to experimentally measurable quantities. Each methylene proton pair is represented by an effective spin-1/2 particle, with basis states corresponding to the central triplet and singlet states:

$$|\uparrow\rangle \equiv |\mathrm{T}_0\rangle = \frac{1}{\sqrt{2}}(|\alpha\beta\rangle + |\beta\alpha\rangle),$$
$$|\downarrow\rangle \equiv |\mathrm{S}_0\rangle = \frac{1}{\sqrt{2}}(|\alpha\beta\rangle - |\beta\alpha\rangle), \quad (7)$$

Here, $\alpha$ and $\beta$ denote the single spin-1/2 particle Zeeman states. A chain of $N$ methylene groups is thereby mapped onto $N$ coupled effective spins.

Under this mapping, the scalar-coupling Hamiltonian projected onto the $\{T_0, S_0\}$ manifold becomes an effective one-dimensional chain of interacting spin-1/2 particles [9]:

$$\hat{H} = J_{intra} \sum_{i=1}^{N}(\hat{I}_{i,z} - \frac{1}{4}\hat{1}) + \sum_{i=1}^{N-1}\frac{\Delta J_i}{2}(\hat{I}_i^+\hat{I}_{i+1}^- + \hat{I}_i^-\hat{I}_{i+1}^+) + \sum_{i=1}^{N-1}\frac{\Delta J_i}{2}(\hat{I}_i^+\hat{I}_{i+1}^+ + \hat{I}_i^-\hat{I}_{i+1}^-). \quad (8)$$

Here, $N$ is the number of sites in the chain, and $J_{intra}$ is the intrapair coupling between the two protons of a $CH_2$ group, assumed to be identical for all sites. The parameter $\Delta J_i$ denotes the difference between the inequivalent interpair scalar couplings connecting sites $i$ and $i + 1$, as defined in Appendix B. The operator $\hat{I}_{i,z}$ is the z component of the effective spin operator at site $i$, $\hat{1}$ is the identity operator, and $\hat{I}_i^+$ and $\hat{I}_i^-$ are the corresponding raising and lowering operators, respectively. The first term in Eq. (8) describes the local energy splitting between the $T_0$ and $S_0$ states. The second describes exchange between neighbouring effective spins, whereas the third simultaneously flips both spins.

The construction of the reduced basis and the derivation of Eq. (8), including the relationship between $\Delta J_i$ and the underlying proton–proton couplings, are detailed in Appendix B.

## B. Three-site and four-site benchmark Hamiltonians

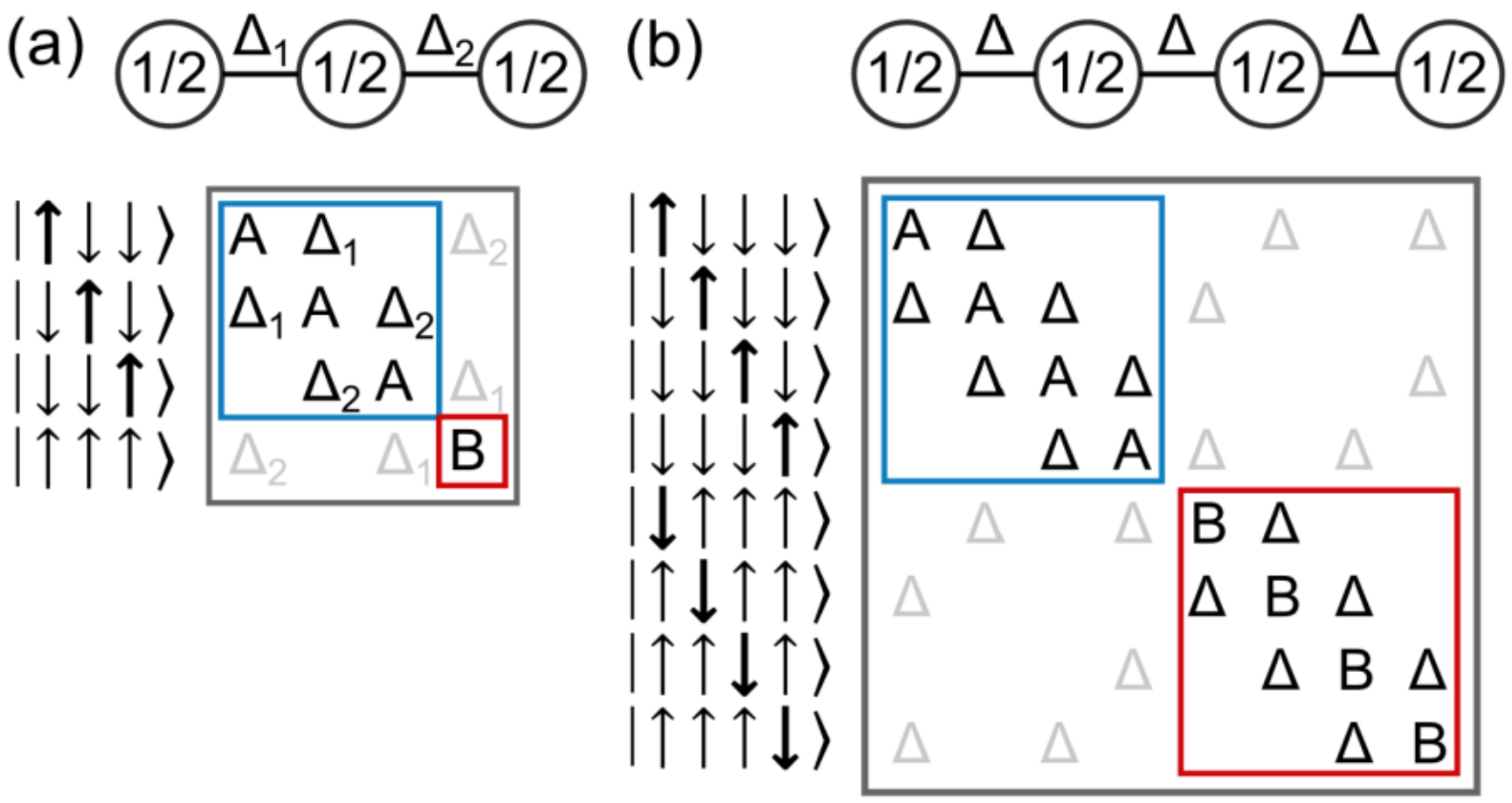

**Fig. 2. Block structure of the Hamiltonian matrices for aliphatic spin chains containing three (a) and four (b) sites.** Blue and red outlines indicate the manifolds with diagonal energies A and B, respectively. Black off-diagonal elements describe exchange between neighbouring sites, whereas grey elements connect the two manifolds through simultaneous neighbouring-site spin flips. For (a) $\Delta_1 = \Delta J_1/2$, $\Delta_2 = \Delta J_2/2$, A = -5/4 $J_{\text{intra}}$, B = 3/4 $J_{\text{intra}}$. For (b), $\Delta = \Delta J/2$, A = -2 $J_{\text{intra}}$, B = 0.

To construct benchmark eigenvalue problems of increasing complexity, we consider spin chains containing three and four sites. The Hamiltonian blocks whose eigenvalue differences determine the frequencies of observable transitions in the corresponding NMR spectra are shown in **Fig. 2** [9]. For the three-site chain, the two nearest-neighbour couplings $\Delta J_1$ and $\Delta J_2$ are treated independently, allowing both the uniform case, $\Delta J_1 = \Delta J_2$, and the nonuniform case, $\Delta J_1 \neq \Delta J_2$, to be examined. For the four-site chain, we consider uniform nearest-neighbour couplings, $\Delta J_1 = \Delta J_2 = \Delta J_3 = \Delta J$. Together, the uniform and nonuniform three-site cases and the uniform four-site case form a controlled progression from an analytically tractable reference problem to more general parameter-dependent Hamiltonians of increasing dimension.

## C. Perturbative reference for the uniform three-site chain

For the uniform three-site chain, defined by $\Delta J_1 = \Delta J_2 = \Delta J$, the four eigenenergies were previously calculated to second order in $\Delta J$ using time-independent perturbation theory for degenerate states [6,8]. The coupling hierarchy defining the perturbative regime is summarized in Appendix B. The resulting expressions provide an independent analytical reference for evaluating the ASN approximations:

$$E_1 = \frac{3}{4}J_{intra} + \frac{1}{4}\frac{\Delta J^2}{J_{intra}}, \tag{9a}$$

$$E_2 = -\frac{5}{4}J_{intra} + \frac{1}{\sqrt{2}}\Delta J - \frac{1}{8}\frac{\Delta J^2}{J_{intra}}, \tag{9b}$$

$$E_3 = -\frac{5}{4}J_{intra}, \tag{9c}$$

$$E_4 = -\frac{5}{4}J_{intra} - \frac{1}{\sqrt{2}}\Delta J - \frac{1}{8}\frac{\Delta J^2}{J_{intra}}. \quad (9d)$$

These expressions contain constant, linear, and quadratic terms in $\Delta J$ and therefore provide a local analytical reference within the perturbative regime. For the nonuniform three-site chain, a corresponding two-parameter perturbative reference can be derived from the same degenerate manifold, as shown in Sec. III.D. The four-site problem then tests whether the ASN can obtain reproducible symbolic approximations for multiple eigenenergy branches of a larger Hamiltonian.

## D. Nonuniform three-site chain: two-parameter perturbative structure

For the nonuniform chain, the threefold-degenerate A manifold is described at first order by the exchange matrix (1/2)[[0, $\Delta J_1$, 0], [$\Delta J_1$, 0, $\Delta J_2$], [0, $\Delta J_2$, 0]]. Its eigenvalues are 0 and $\pm(1/2)\sqrt{\Delta J_1^2 + \Delta J_2^2}$. The leading splitting therefore depends on the radial coordinate in the two-parameter coupling plane and is nonpolynomial at the degeneracy. Defining $S = \Delta J_1^2 + \Delta J_2^2$ and evaluating the second-order corrections in the eigenbasis of this first-order matrix gives:

$$E_1 = \frac{3}{4}J_{intra} + \frac{1}{8}\frac{S}{J_{intra}} + O\left(\frac{S^{3/2}}{J_{intra}^2}\right), \quad (10a)$$

$$E_2 = -\frac{5}{4}J_{intra} + \frac{1}{2}\sqrt{S} - \frac{1}{4}\frac{\Delta J_1^2 \Delta J_2^2}{J_{intra}S} + O\left(\frac{S^{3/2}}{J_{intra}^2}\right), \quad (10b)$$

$$E_3 = -\frac{5}{4}J_{intra} - \frac{1}{8}\frac{\left(\Delta J_2^2 - \Delta J_1^2\right)^2}{J_{intra}S} + O\left(\frac{S^{3/2}}{J_{intra}^2}\right), \quad (10c)$$

$$E_4 = -\frac{5}{4}J_{intra} - \frac{1}{2}\sqrt{S} - \frac{1}{4}\frac{\Delta J_1^2 \Delta J_2^2}{J_{intra}S} + O\left(\frac{S^{3/2}}{J_{intra}^2}\right). \quad (10d)$$

The derivation is given in Appendix D. At $S = 0$, the rational terms containing $1/S$ are defined by continuity and assigned their zero-coupling limit of zero. Along the positive uniform line $\Delta J_1 = \Delta J_2 = \Delta J \geq 0$, Eqs. (10a) - (10d) reduce to Eqs. (9a) - (9d) through second order. On a 101 × 101 grid covering $0 \leq \Delta J_1/J_{intra}$, $\Delta J_2/J_{intra} \leq 0.2$, the corresponding dimensionless eigenenergies reproduce exact diagonalization with a root-mean-square error (RMSE) of $4.28 \times 10^{-5}$ and a maximum absolute error of $2.64 \times 10^{-4}$.

# IV. Preparation of data for symbolic regression

## A. Dimensionless formulation of the eigenvalue problem

Direct symbolic regression of the eigenenergies $E_i(J_{intra}, \Delta J_1, \Delta J_2)$ proved inefficient because the learned expressions generally violated dimensional consistency. Since both the Hamiltonian parameters and the eigenenergies are measured in units of frequency, physically meaningful analytical expressions must preserve the corresponding dimensionality. Since the Hamiltonian in Eq. (8) is linear in $J_{intra}$ and $\Delta J_i$, multiplication of all coupling parameters by a common factor produces the same scaling of every eigenenergy. Because $J_{intra}$ sets the overall energy scale, the three-site eigenvalue problem can be expressed in terms of two independent dimensionless coupling ratios, which we define as follows:

$$r_i \equiv \frac{\Delta J_i}{J_{intra}},\ i = 1{,}2;\ \varepsilon_k \equiv \frac{E_k}{J_{intra}},\ k = 1, \ldots, 4. \tag{11}$$

This normalization preserves dimensional consistency and transforms the original mapping $E_k(J_{intra}, \Delta J_1, \Delta J_2)$ into a two-dimensional regression problem $\varepsilon_k(r_1, r_2)$.

## B. Assignment and symmetry of the dimensionless eigenenergy surfaces

A second challenge concerns the consistent assignment of the eigenenergy surfaces. For each pair $(r_1, r_2)$, the Hamiltonian in Eq. (8) was diagonalized numerically. The resulting surfaces were labelled according to the convention $\varepsilon_1 > \varepsilon_2 \geq \varepsilon_3 \geq \varepsilon_4$, consistent with **Fig. 3** and Eqs. (9) after division by $J_{intra}$. Because physically $J_{intra} < 0$ [8,11,12], this ordering is the reverse of that of the signed physical eigenenergies, which satisfy $E_1 < E_2 \leq E_3 \leq E_4$. The adopted convention defines four single-valued dimensionless eigenenergy surfaces over the two-parameter space.

Along the uniform-chain line, $r_1 = r_2 = r$, the branches $\varepsilon_2$, $\varepsilon_3$, and $\varepsilon_4$ become degenerate at $r = 0$. If the normalized perturbative branches obtained from Eqs. (9) are continued from $r > 0$ to $r < 0$, the branches corresponding to $E_2$ and $E_4$ exchange their positions in the dimensionless-energy ordering at the origin. Maintaining fixed order-based labels therefore replaces the leading linear dependence on r with a dependence on $|r|$. The resulting ordered surfaces remain continuous but are not differentiable at the degeneracy. Away from this point, as illustrated in **Fig. 3(c)**, all four surfaces vary smoothly.

The ordered eigenenergy surfaces exhibit symmetry properties that can be exploited to reduce the training domain. In particular, they are invariant under independent reversal of the sign of either coupling difference and under interchange of $\Delta J_1$ and $\Delta J_2$. The sign-reversal symmetry allows the ASN to be trained within the first quadrant, defined by $0 \leq r_1 \leq 2$ and $0 \leq r_2 \leq 2$, with the remaining quadrants subsequently reconstructed by reflection. The upper boundary of 2 was chosen to include sufficiently pronounced nonlinear behaviour while retaining a parameter range over which, along the uniform-chain line, the second-order perturbative expressions remain relatively close to the numerically calculated eigenenergies (**Fig. 3(b)**). This restriction places the non-differentiable point at the corner of the training domain and avoids requiring a polynomial network to reproduce the cusp across positive and negative coupling values. Because such non-differentiable behaviour cannot be represented exactly by a finite polynomial expansion, it would otherwise present an additional challenge for symbolic regression based only on polynomial activation functions.

Restricting training to the first quadrant removes the sign-change cusp across positive and negative couplings, but it does not make the ordered spectrum analytic at the origin. Equation (10b) shows explicitly that the leading splitting contains $\sqrt{\Delta J_1^2 + \Delta J_2^2}$. A finite polynomial can therefore approximate but cannot exactly reproduce the local two-dimensional spectral geometry at the degeneracy.

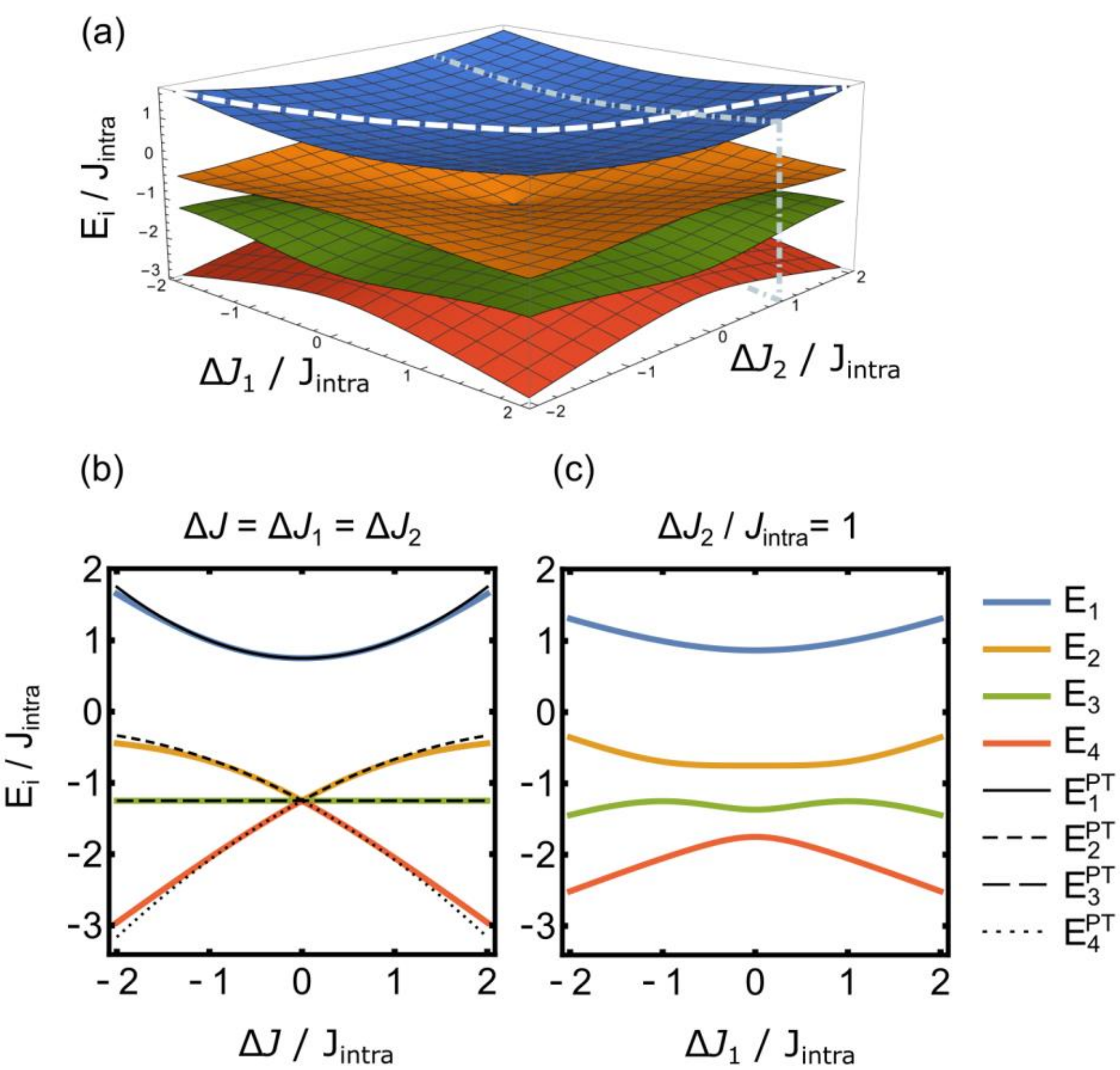

**Fig. 3. Dimensionless eigenenergy surfaces and representative cross-sections for the Hamiltonian of the three-site spin chain**. All energies and coupling constants are scaled by $J_{intra}$. (a) Numerically calculated eigenenergy surfaces $E_i/J_{intra}$ as functions of the dimensionless couplings $\Delta J_1/J_{intra}$ and $\Delta J_2/J_{intra}$. The white dashed and the grey dot-dash lines indicate cross-sections used for (b) and (c), respectively. (b) Cross-section corresponding to the uniform chain, $\Delta J_1 = \Delta J_2 = \Delta J$. Coloured curves show the numerically calculated, energy-ordered eigenenergies, while the black solid, dashed, dash-dotted, and dotted curves show the second-order perturbative expressions in Eqs. (9). Across $\Delta J = 0$, the perturbative branches corresponding to $E_2$ and $E_4$ exchange their energy ordering. Maintaining a fixed ordering therefore introduces non-differentiability at the degeneracy point. (c) Cross-section for fixed $\Delta J_2/J_{intra} = 1$, illustrating the smooth variation of the eigenenergies away from the degeneracy. In the full two-dimensional parameter space, the eigenenergy surfaces remain ordered and should therefore be labelled according to their energies, $\varepsilon_1 > \varepsilon_2 \geq \varepsilon_3 \geq \varepsilon_4$, rather than by continuing the one-dimensional perturbative branches through $\Delta J = 0$.

# V. Symbolic regression of spin-chain eigenenergies

## A. Three-site spin chain

Having defined the dimensionless coupling ratios $r_1$ and $r_2$ and the ordered eigenenergy surfaces $\varepsilon_i$, we applied the ASN to the simultaneous symbolic regression of the four eigenenergy branches of the three-site chain. Candidate architectures with different depths and widths were compared in terms of predictive error and symbolic complexity, quantified by the total number of operations in the extracted expressions. The architecture (2, 16, 4), comprising two input nodes, one hidden layer with 16 nodes, and four output nodes, was retained as a compact model balancing predictive accuracy and expression complexity.

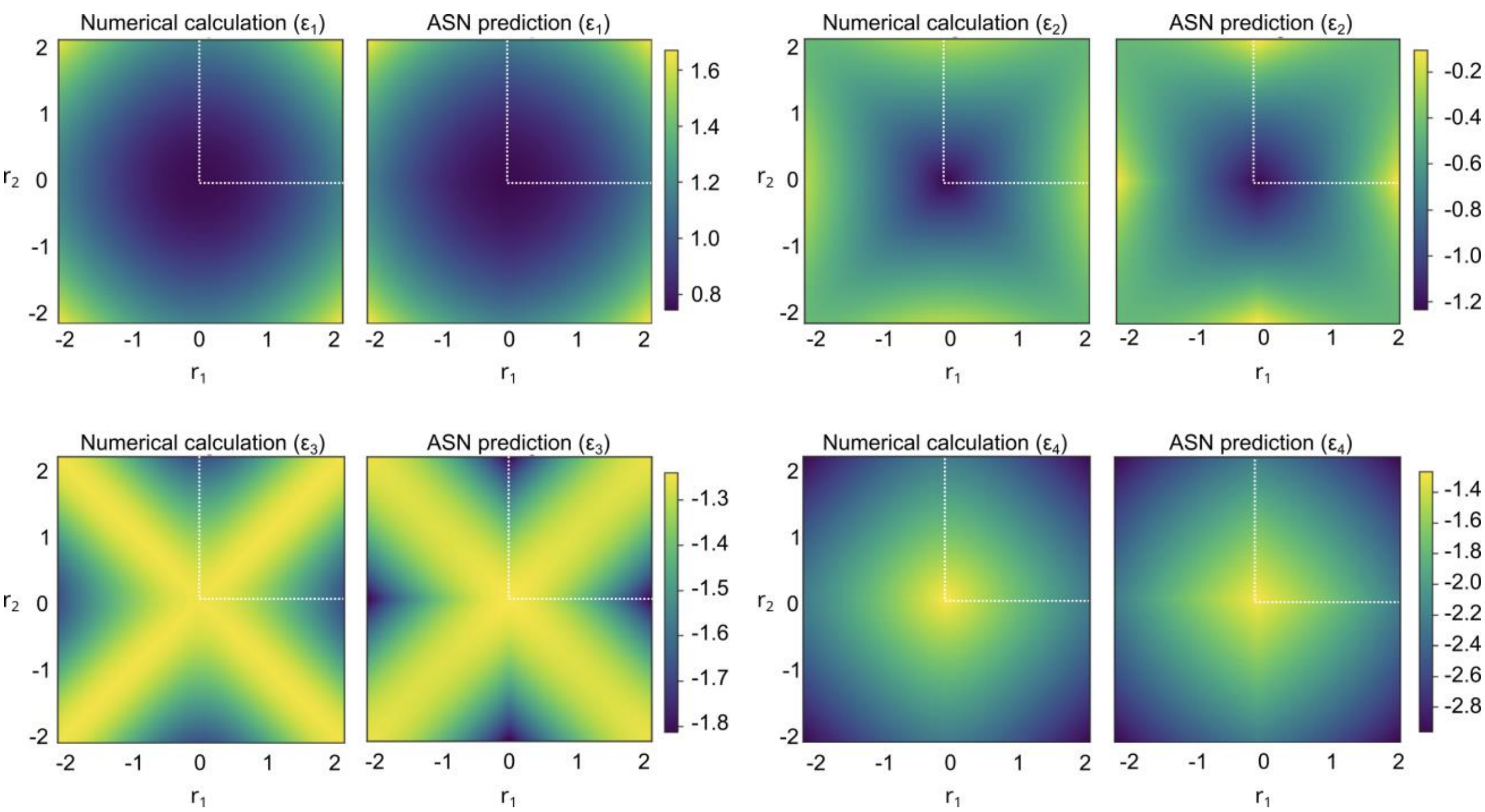

**Fig. 4. Comparison of numerically calculated and ASN-predicted dimensionless eigenenergy surfaces for the three-site spin chain**. Here, $r_i = \Delta J_i/J_{intra}$ and $\varepsilon_k = E_k/J_{intra}$. For each $\varepsilon_k$, the left and right panels show the numerical eigenenergies and corresponding ASN predictions, respectively. The ASN was trained within the first-quadrant domain $0 \le r_1, r_2 \le 2$, indicated by the white dotted lines, and the remaining quadrants were reconstructed using the independent sign-reversal symmetries of $r_1$ and $r_2$. A common colour scale is used for the numerical and predicted surfaces within each pair. The activation set used for training was $\{0, x, x^2\}$.

The selected ASN was trained to predict the four dimensionless eigenenergies $\varepsilon_i$ simultaneously using the activation set $\{0,\ x,\ x^2\}$. Inclusion of $|x|$ among the candidate activations was also examined but did not improve the prediction accuracy and was therefore omitted from the final activation set. The complete training procedure and hyperparameters are described in Methods.

**Fig. 4** compares the numerically calculated dimensionless eigenenergy surfaces with the ASN predictions. Within the first-quadrant training domain, the ASN reproduces the curvature and overall topology of all four surfaces. Predictions in the remaining quadrants were obtained by reflection rather than by evaluating the polynomial network outside its training domain. Consequently, the cusp-like and cross-shaped structures in the full-domain representation arise from combining the first-quadrant approximation with the exact sign-reversal symmetries of the Hamiltonian. They should not be interpreted as nonanalytic behavior learned directly by the polynomial network.

For the selected ASN, the original training run yielded an overall mean-squared error of $3.20 \times 10^{-4}$ across the complete sampled data set and all four outputs, corresponding to a root-mean-square error of $1.79 \times 10^{-2}$. The coefficients of determination were $R^2$ = 0.9994, 0.9888, 0.9483, and 0.9979 for $\varepsilon_1$–$\varepsilon_4$, respectively. These metrics were calculated from the numerical network outputs before coefficient thresholding, symmetry averaging, and rounding.

To obtain an interpretable representation, the trained ASN was converted into symbolic form, and the resulting expressions for $\varepsilon_i$ were expanded in the dimensionless inputs $r_1$ and $r_2$. Multiplication by the signed $J_{\text{intra}}$ then restored physical dimensions, $E_i^{ASN} = J_{intra}\varepsilon_i^{ASN}$, with the branch labels inherited from the dimensionless-energy ordering defined above. Because the interchange symmetry $r_1 \leftrightarrow r_2$ was not imposed explicitly during training, symmetry-equivalent coefficients were averaged after extraction and rounded to three decimal places. The expressions below therefore result from direct algebraic extraction followed by two explicit post-processing steps: symmetry averaging and coefficient rounding.

$$E_1^{ASN} = 0.743J_{intra} + 0.022(\Delta J_1 + \Delta J_2) + 0.094\frac{(\Delta J_1)^2+(\Delta J_2)^2}{J_{intra}} + 0.023\frac{\Delta J_1 \Delta J_2}{J_{intra}}, \tag{12a}$$

$$E_2^{ASN} = -1.233J_{intra} + 0.361(\Delta J_1 + \Delta J_2) + 0.104\frac{(\Delta J_1)^2+(\Delta J_2)^2}{J_{intra}} - 0.372\frac{\Delta J_1 \Delta J_2}{J_{intra}}, \tag{12b}$$

$$E_3^{ASN} = -1.239J_{intra} - 0.028(\Delta J_1 + \Delta J_2) - 0.133\frac{(\Delta J_1)^2+(\Delta J_2)^2}{J_{intra}} + 0.288\frac{\Delta J_1 \Delta J_2}{J_{intra}}, \tag{12c}$$

$$E_4^{ASN} = -1.272J_{intra} - 0.355(\Delta J_1 + \Delta J_2) - 0.062\frac{(\Delta J_1)^2+(\Delta J_2)^2}{J_{intra}} + 0.054\frac{\Delta J_1 \Delta J_2}{J_{intra}}. \tag{12d}$$

Equations (12) provide compact global quadratic approximations to the dependence of the four eigenenergy branches on $r_1$ and $r_2$ within the first-quadrant training domain, $0 \le r_1, r_2 \le 2$. The remaining quadrants are reconstructed using the independent sign-reversal symmetries. Unlike the local uniform-chain expressions in Eqs. (9a)–(9d), Eqs. (12a)–(12d) apply to nonuniform chains over a broad finite domain. Unlike the two-parameter perturbative expressions in Eqs. (10a)–(10d), however, their polynomial form does not reproduce the exact radial nonanalyticity at the degeneracy.

Along the uniform-chain line ($r_1 = r_2 = r$), the ASN expressions can be compared directly with the independently derived perturbative results (Eqs. (9)). The ASN reproduces their qualitative structure, including the nearly constant $\varepsilon_3$ branch and approximately opposite linear coefficients for $\varepsilon_2$ and $\varepsilon_4$. Specifically, the ASN coefficients of (+0.722) and (-0.710), respectively, are close to the perturbative values $+1/\sqrt{2}$ and $-1/\sqrt{2}$. The residual differences between the two sets of coefficients are expected because perturbation theory provides a local expansion about $r = 0$, whereas the ASN was trained to approximate all four eigenenergy surfaces over the full two-dimensional domain $(r_1, r_2)$. The extracted ASN coefficients therefore describe global symbolic approximations and should not be interpreted as perturbative coefficients.

Because the ASN was trained directly on numerically calculated eigenenergies, its accuracy within the sampled domain is not constrained by the small-parameter requirement of perturbation theory. Along the uniform-chain line $0 \leq r \leq 2$, the rounded expressions in Eqs. (12a)–(12d) have an RMSE of $1.4 \times 10^{-2}$ over the four eigenenergy branches, compared with $5.0 \times 10^{-2}$ for the second-order perturbative expressions. Perturbation theory nevertheless remains substantially more accurate close to r = 0 and reproduces the constant $\varepsilon_3$ branch exactly. On an independent 121 × 121 grid covering the complete first-quadrant domain, the rounded ASN expressions yield an MSE of $3.62 \times 10^{-4}$ and an RMSE of $1.90 \times 10^{-2}$, compared with an RMSE of $1.79 \times 10^{-2}$ for the unrounded network output. This small increase quantifies the loss of accuracy introduced by symbolic post-processing. All reported comparisons are restricted to the training domain because the behaviour of the extracted expressions outside this range has not been established.

These results show that the ASN can extract compact global analytical approximations from numerical eigenenergy data. Predictive accuracy alone, however, does not establish an advantage of learned activation selection over direct polynomial regression. We therefore next compare the ASN expressions with fixed-basis least-squares polynomial fits.

## B. Fixed-basis and physics-informed baselines

To determine whether the ASN provides an advantage over direct polynomial regression for this benchmark, we fitted ordinary least-squares models to the same 100,000 dimensionless input pairs. The symmetry-constrained quadratic basis $\{1,\ r_1 + r_2,\ r_1^2 + r_2^2,\ r_1 r_2\}$ gives a complete-sample RMSE of $1.62 \times 10^{-2}$, slightly below the value of $1.79 \times 10^{-2}$ obtained for the unrounded ASN output. The fitted coefficients closely match those extracted from the ASN in Eqs. (12a)–(12d), showing that the network learns a global quadratic approximation that can also be obtained by direct linear regression once the appropriate basis has been specified.

Adding the physics-informed radial feature $\rho = \sqrt{r_1^2 + r_2^2}$ reduces the held-out RMSE to $1.40 \times 10^{-2}$. The improvement is particularly pronounced near the degeneracy: over $0 \leq r_1, r_2 \leq 0.2$, the radial-feature model gives an RMSE of $6.58 \times 10^{-5}$, compared with $1.74 \times 10^{-3}$ for the symmetry-constrained quadratic model. The perturbative expressions in Eqs. (10a)–(10d) remain the most accurate in this local region, with an RMSE of $4.28 \times 10^{-5}$.

These comparisons clarify the role of the ASN in this benchmark. The network produces compact symbolic approximations, but it does not outperform ordinary least-squares regression when the appropriate polynomial basis is known in advance. Its potential advantage lies instead in making the selection and composition of analytical building blocks part of the training procedure when several functional forms are plausible and the appropriate basis is not known beforehand.

## C. Four-site spin chain

To extend the analysis beyond the three-site benchmark, we applied the ASN to the four-site chain shown in **Fig. 2(b)**, with all nearest-neighbor coupling differences set equal to $\Delta J$. Using $J_{intra}$ as the energy scale, the eight dimensionless eigenenergies $\varepsilon_k = E_k/J_{intra}$ depend on the single dimensionless input $r = \Delta J/J_{intra}$. At $r = 0$, the eigenenergies form two fourfold-degenerate manifolds at $\varepsilon = 0$ and $\varepsilon = -2$. For nonzero $r$, the exchange terms split the levels within each manifold, while the double-quantum terms couple the two manifolds, producing a more complex eigenenergy structure than in the three-site benchmark [9]. An ASN with architecture (1, 10, 8), comprising one input, a single hidden layer with 10 nodes, and eight outputs, was trained to predict all eight eigenenergy branches simultaneously using the same zero, identity, and quadratic activation functions as in the three-site analysis.

**Fig. 5** compares the numerically calculated eigenenergies with the ASN predictions. The network reproduces all eight branches over the sampled range $0 \leq r \leq 3$. The held-out validation mean-squared error was $6.2 \times 10^{-5}$, corresponding to an RMS deviation of $7.9 \times 10^{-3}$. When evaluated over the complete sampled dataset, the per-branch coefficients of determination ranged from $R^2 = 0.9918$ to $0.9998$. The largest deviations occur near the upper boundary of the sampled range, where the exact eigenenergies depart most strongly from quadratic behaviour.

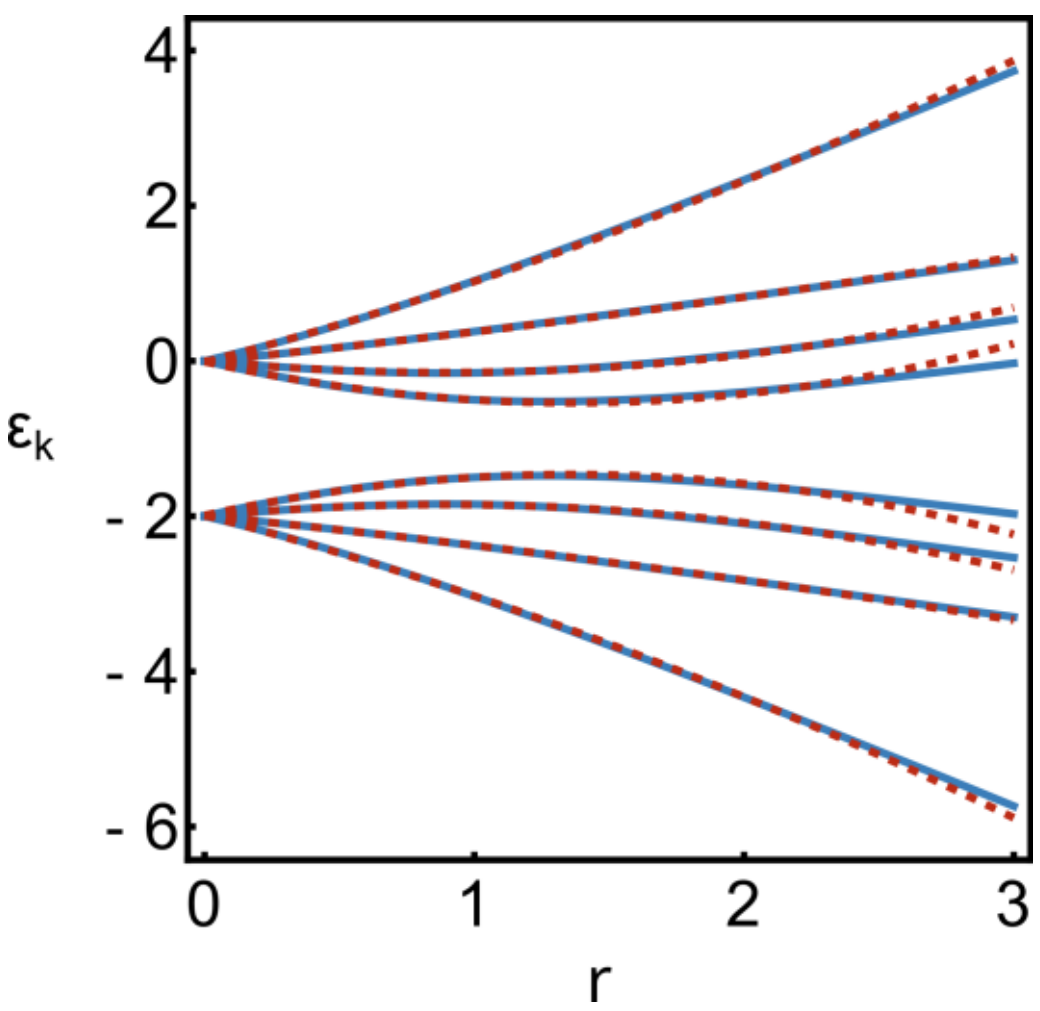

**Fig. 5. Numerical and ASN-predicted eigenenergy branches for the four-site spin chain with uniform nearest-neighbour couplings.** The dimensionless eigenenergies $\varepsilon_k = E_k/J_{intra}$ are shown as functions of $r = \Delta J/J_{intra}$. Results obtained by numerical diagonalization are shown as blue solid curves, and the corresponding ASN predictions are shown as red dashed curves. At $r = 0$, the eight branches form two fourfold-degenerate manifolds at $\varepsilon = 0$ and $\varepsilon = -2$.

After activation selection and symbolic extraction, every predicted eigenenergy was reduced to a quadratic polynomial containing at most three terms (Eqs. (13)). The resulting expressions have the general form $E_k/J_{intra} = c_{k,0} + c_{k,1}r + c_{k,2}r^2$. Multiplication by the signed $J_{intra}$ restored the physical dimensions while retaining the branch labels defined by the descending dimensionless-energy ordering shown in **Fig. 5**. The resulting analytical approximations are:

$$E_1^{ASN} = -0.016J_{intra} + 0.915\Delta J + 0.128\Delta J^2/J_{intra}, \tag{13a}$$
$$E_2^{ASN} = -0.006J_{intra} + 0.351\Delta J + 0.032\Delta J^2/J_{intra}, \tag{13b}$$
$$E_3^{ASN} = -0.337\Delta J + 0.188\Delta J^2/J_{intra}, \tag{13c}$$
$$E_4^{ASN} = -0.010J_{intra} - 0.772\Delta J + 0.282\Delta J^2/J_{intra}, \tag{13d}$$
$$E_5^{ASN} = -1.991J_{intra} + 0.775\Delta J - 0.284\Delta J^2/J_{intra}, \tag{13e}$$

$$E_6^{ASN} = -2.001 J_{intra} + 0.338 \Delta J - 0.188 \Delta J^2 / J_{intra}, \quad (13f)$$
$$E_7^{ASN} = -1.995 J_{intra} - 0.349 \Delta J - 0.034 \Delta J^2 / J_{intra}, \quad (13g)$$
$$E_8^{ASN} = -1.979 J_{intra} - 0.933 \Delta J - 0.118 \Delta J^2 / J_{intra}. \quad (13h)$$

The coefficients in Eqs. (13a)– (13h) reproduce the qualitative perturbative structure of the four-site Hamiltonian. At $\Delta J = 0$, the constant terms recover the two fourfold-degenerate manifolds at $E_k/J_{intra} = 0$ and $E_k/J_{intra} = -2$, with deviations not exceeding 0.021. To first order in $\Delta J$, diagonalization of either tridiagonal exchange block gives the linear coefficients $\cos(k\pi/5)$, with $k \in \{1, \ldots, 4\}$, corresponding approximately to ±0.809 and ±0.309. The coefficients multiplying $\Delta J$ in Eqs. (13a)–(13h) follow the same sequence of signs and magnitudes but do not coincide exactly with these local perturbative values because the ASN minimizes the global approximation error over the complete sampled interval $0 \leq \Delta J/J_{intra} \leq 3$. The terms proportional to $\Delta J^2/J_{intra}$ describe the curvature generated by double-quantum coupling between the two manifolds.

The exact eigenenergy set is symmetric about $E_k/J_{intra} = -1$, such that each pair of branches $E_k$ and $E_{9-k}$ sums to $-2J_{intra}$ for $k = 1, \ldots, 4$. Although this constraint was not imposed during training, Eqs. (13a)–(13h) reproduce it with a maximum deviation of approximately $0.041|J_{intra}|$ over the sampled interval. A symmetry-constrained ASN could enforce the relation exactly by predicting four branches and constructing the remaining four algebraically.

## D. Independent verification and quadratic baseline

To provide a controlled comparison with fixed-basis quadratic regression, we repeated the four-site training using an independently written ASN implementation and evaluated both approaches on the same 10,000-parameter sample and 80/20 train-test split. The network architecture and activation set were unchanged from those described in Sec. V C, and the complete training settings are given in Methods. The independent run yielded a held-out MSE of $5.67 \times 10^{-5}$ and an RMSE of $7.53 \times 10^{-3}$, close to the corresponding values of $6.2 \times 10^{-5}$ and $7.9 \times 10^{-3}$ obtained in the primary calculation. The small difference reflects the use of a separate training and evaluation run. The coefficient of determination $R^2$ ranged from 0.9935 to 0.9998 across the eight branches, and all 11 activation-selection coefficient vectors had a support size of one after the transition to sparsemax.

On the same parameter sample and train-test split, ordinary least-squares regression using the quadratic basis $\{1, r, r^2\}$ gives a slightly lower MSE of $4.93 \times 10^{-5}$ and an RMSE of $7.02 \times 10^{-3}$. All errors refer to the dimensionless eigenenergies $E_k/J_{intra}$. Representative fitted expressions are

$$E_1^{LS} = -0.014 J_{intra} + 0.908 \Delta J + 0.133 \Delta J^2 / J_{intra}, \quad (14a)$$
$$E_2^{LS} = -0.005 J_{intra} + 0.349 \Delta J + 0.034 \Delta J^2 / J_{intra}, \quad (14b)$$

which closely match the corresponding ASN expressions in Eqs. (13a) and (13b). The independent calculation therefore confirms both the numerical accuracy of the four-site ASN and the emergence of one-sparse activation selections. It does not, however, provide a predictive advantage over regression with a fixed quadratic basis. The four-site problem nevertheless provides a nontrivial multi-output test of reproducibility and demonstrates that the learned activation selections can collapse to a single analytical function at every node.

# VI. Practical considerations and limitations

The ASN architecture should be chosen according to the expected complexity of the target expression. Network width provides independently weighted intermediate variables, whereas depth increases compositional complexity. With the candidate set {0, $x$, $x^2$}, the maximum polynomial degree can grow as $2^L$ across L consecutive ASN layers, while mixed terms arise from applying quadratic activations to linear combinations of several variables.

When a suitable low-order polynomial basis is known in advance, ordinary least-squares regression provides the natural baseline: it is faster, globally optimized, and immediately symbolic. The ASN is most useful when the appropriate analytical basis is uncertain and must be selected from several candidate functions. The choice of this library determines the class of expressions that can be recovered; domain-sensitive or singular functions require particularly careful definition, optimization, and validation.

The extracted expressions should be regarded as approximations within their specified training domains, because extrapolation is not guaranteed. Branch assignment near degeneracies and avoided crossings must be treated consistently, and stability should be assessed across multiple training seeds. Exact spectral constraints should also be imposed when available. For example, the four three-site eigenenergies sum to $-3J_{intra}$, while paired four-site eigenenergies sum to $-2J_{intra}$. Enforcing these relations would reduce the number of independent outputs and ensure that the extracted expressions preserve the corresponding Hamiltonian symmetries.

# VII. Conclusions

We introduced the activation-selection network as a differentiable symbolic-regression framework in which each node learns a sparse selection from a predefined library of analytical functions. The resulting expressions can be extracted directly from the trained network. Dimensional normalization preserves the physical scaling of the problem, while compositions of the candidate functions {0, $x$, $x^2$} generate mixed polynomial terms and progressively higher degrees with increasing network depth.

For the three-site spin chain, the ASN produced compact global quadratic approximations to the four dimensionless eigenenergy surfaces. Along the uniform-chain line, these expressions reproduce the qualitative structure of second-order perturbation theory. For the nonuniform chain, the two-parameter perturbative analysis showed that the exact local splitting contains the radial dependence $\sqrt{\Delta J_1^2 + \Delta J_2^2}$, which lies outside the polynomial function library used here. For the four-site chain, the ASN represented all eight eigenenergy branches simultaneously. An independently written implementation reproduced the numerical accuracy of the original calculation and yielded a support size of one for every activation-selection coefficient vector after the transition to sparsemax.

Fixed-basis quadratic least-squares regression was nevertheless slightly more accurate for both benchmarks when the appropriate polynomial basis was specified in advance. The central contribution of the ASN is therefore not a predictive advantage over regression with a known basis, but the ability to make the selection and composition of candidate analytical functions trainable and to extract compact global approximations directly from numerical data. Stronger evidence for adaptive function selection will require benchmarks with competing nonpolynomial primitives, multiple training seeds, structured extrapolation tests, symmetry-constrained outputs, and direct comparisons with established symbolic-regression methods.

# VIII. Methods

## A. Numerical dataset generation

For the three-site chain, the dimensionless coupling ratios $r_1 = \Delta J_1/J_{intra}$ and $r_2 = \Delta J_2/J_{intra}$ defined in Eq. (11) were sampled independently and uniformly over $0 \le r_1, r_2 \le 2$. A total of 100,000 parameter pairs were generated using NumPy random seed 42. For each pair, the 4 × 4 Hamiltonian block shown in **Fig. 2(a)** was diagonalized directly. The eigenvalues were divided by the signed $J_{intra}$ and ordered according to $\varepsilon_1 > \varepsilon_2 \ge \varepsilon_3 \ge \varepsilon_4$, as defined in Sec. IV B. Because $J_{intra} < 0$ for the physical systems considered here, this is the reverse of the ordering of the signed dimensional eigenenergies.

For the primary four-site dataset, $J_{intra}$ and $\Delta J$ were sampled independently over their respective intervals. Consequently, the derived dimensionless ratio $r = \Delta J/J_{intra}$ followed a nonuniform distribution spanning approximately $0 \le r \le 3$. Numerical eigenenergies were first calculated on a 200 × 200 grid spanning $-32 \le J_{intra} \le -5$ and $-15 \le \Delta J \le -0.1$ in frequency units. Continuous eigenenergy branches were tracked between neighbouring grid points by maximizing the overlap of the corresponding eigenvectors. Linear interpolation of the resulting branch surfaces was then used to generate 10,000 examples with $J_{intra}$ and $\Delta J$ sampled independently over these intervals. After normalization, the eight outputs $\varepsilon_k = E_k/J_{intra}$ were ordered from highest to lowest.

The independent four-site verification used a distinct dataset, described in Sec. VIII E, for which the 8 × 8 Hamiltonian was diagonalized directly at every sampled value. The primary and independent calculations therefore constitute separate training and evaluation runs rather than repeated evaluations of an identical fitted model.

## B. Network training and architecture selection

All ASN models were implemented in Python using PyTorch. Each ASN layer used the candidate activation set {0, $x$, $x^2$} and included a linear bias. During the first 90% of the training epochs, activation-selection coefficients were calculated using softmax Eq. (1). The $L_{0.5}$-type term in Eq. (3), multiplied by a coefficient of 0.1, was added to the fitting loss to encourage coefficient vectors close to a simplex vertex. During the final 10% of training, softmax was replaced by sparsemax Eq. (4) and the regularization term was removed. Sparsemax permits exact zeros but does not itself guarantee a support size of one.

For the three-site problem, candidate depths and widths were compared using prediction error and symbolic complexity, defined as the total number of arithmetic operations in the extracted expressions. The non-dominated candidates formed a Pareto front, from which the architecture (2, 16, 4) was selected as a compact model. Here and below, an architecture tuple lists the numbers of input, hidden, and output nodes. The selected model was trained for 1000 epochs using the Adam optimizer, a learning rate of $10^{-3}$, a batch size of 4096, and mean absolute error as the primary fitting loss. Further details of the Pareto selection and the polynomial structure generated by this architecture are given in Appendix C.

For the primary four-site calculation, the architecture was (1, 10, 8). The model was trained for 5000 epochs using Adam with a learning rate of $10^{-3}$, a batch size of 8000, and mean-squared error as the primary fitting loss. For both primary benchmarks, the data were divided randomly into 80% training and 20% validation subsets using PyTorch random seed 42.

## C. Symbolic extraction and model evaluation

After training, the activation-selection coefficients, linear weights, and biases were converted directly into symbolic expressions. Consecutive ASN layers were composed algebraically, and the resulting expressions were expanded and simplified using SymPy. Coefficients with absolute values below $10^{-3}$ were removed, and the remaining coefficients were rounded to three decimal places.

For the three-site chain, coefficients multiplying symmetry-equivalent terms were averaged after extraction because the interchange symmetry $r_1 \leftrightarrow r_2$ was not imposed explicitly during training. Multiplication of the dimensionless expressions by the signed $J_{intra}$ restored physical dimensions and produced Eqs. (12a)–(12d). The same dimensional restoration was applied to the four-site expressions in Eqs. (13a)–(13h). Thus, the published formulas include two explicit post-processing operations: symmetry averaging where applicable and coefficient thresholding and rounding.

Prediction accuracy was quantified using mean-squared error (MSE), root-mean-squared error (RMSE), and the coefficient of determination $R^2$. MSE and RMSE were averaged over all evaluated samples and output branches, whereas $R^2$ was calculated separately for each branch. Metrics for the original neural runs were evaluated before symbolic thresholding and rounding.

Rounded analytical expressions and independently reproduced neural models were evaluated separately so that training error and symbolic post-processing error were not conflated. The support size of an activation-selection coefficient vector was defined as the number of nonzero coefficients after the transition to sparsemax.

## D. Fixed-basis regression baselines

The three-site fixed-basis models used the same 100,000 dimensionless input pairs and the same 80/20 split as the corresponding ASN calculation. Each eigenenergy surface was fitted independently by ordinary least squares. The symmetry-constrained quadratic basis was {1, $r_1 + r_2$, $r_1^2 + r_2^2$, $r_1 r_2$}. The physics-informed basis additionally contained the radial feature $\rho = \sqrt{r_1^2 + r_2^2}$. Coefficients were determined without regularization. Fits to the complete dataset were used to compare coefficients with Eqs. (12a)–(12d), whereas held-out errors were calculated on the validation subset.

To examine the local behavior near the degeneracy, the globally fitted three-site models were evaluated without refitting on a 101 × 101 grid over $0 \leq r_1, r_2 \leq 0.2$. The perturbative expressions in Eqs. (10a)–(10d) were evaluated on the same grid.

For the four-site baseline, each of the eight dimensionless eigenenergy branches was fitted independently using the basis {1, $r$, $r^2$}. The least-squares models, the independently trained ASN, and the rounded expressions in Eqs. (13a)–(13h) were evaluated using the same parameter sample and train-test split. Representative least-squares expressions are given in Eqs. (14a) and (14b).

## E. Independent verification runs

The reduced three-site verification used 20,000 independently generated samples, architecture (2, 16, 4), 400 training epochs, mean absolute error, a learning rate of $10^{-3}$, a batch size of 4096, and the same regularization coefficient and softmax-to-sparsemax schedule as the primary calculation. This reduced run was used only to test numerical reproducibility and the final support sizes; it was not used to obtain Eqs. (12a)–(12d) or the primary three-site error values.

For the independent four-site verification, 10,000 values of $r$ were sampled uniformly over $0 \leq r \leq 3$ using NumPy random seed 42. The eight dimensionless target eigenenergies were obtained by direct diagonalization of the 8 × 8 Hamiltonian at each sampled value, rather than by interpolation of a precomputed surface. The data were divided using an 80/20 split generated with PyTorch random seed 42. The ASN architecture and optimization settings were otherwise those specified for the four-site model in Sec. VIII B. The quadratic baseline was trained and evaluated on the same split. This controlled protocol produced the values reported in Sec. V D.

# Appendices

## Appendix A: Mathematical details of the activation-selection network and multiplication example

**Figure 1(b)** illustrates a two-layer ASN that multiplies two arbitrary inputs $x_1$ and $x_2$ using the candidate set {0, $x$, $x^2$}. The zero activation suppresses unused paths. In the first layer, both input nodes select the identity activation. A convenient set of logits and the resulting sparse coefficient vectors is:

$$\boldsymbol{\alpha}^{(1)} = \begin{bmatrix} -50 & 50 & -50 \\ -50 & 50 & -50 \end{bmatrix}. \tag{A1}$$

$$\mathbf{a}^{(1)} = \begin{bmatrix} 0 & 1 & 0 \\ 0 & 1 & 0 \end{bmatrix}, \qquad v_i^{(1)} = x_i. \tag{A2}$$

The large positive and negative logits are illustrative finite values whose sparsemax projections select the indicated functions. The corresponding weights and biases are

$$\mathbf{w}^{(1)} = \begin{bmatrix} 1 & 1 \\ 1 & -1 \end{bmatrix}, \qquad \mathbf{b}^{(1)} = \begin{bmatrix} 0 \\ 0 \end{bmatrix}. \tag{A3}$$

and therefore

$$y_1 = x_1 + x_2, \qquad y_2 = x_1 - x_2. \tag{A4}$$

In the second layer, both intermediate nodes select the quadratic activation,

$$\boldsymbol{\alpha}^{(2)} = \begin{bmatrix} -50 & -50 & 50 \\ -50 & -50 & 50 \end{bmatrix}. \tag{A5}$$

$$\mathbf{a}^{(2)} = \begin{bmatrix} 0 & 0 & 1 \\ 0 & 0 & 1 \end{bmatrix}, \qquad v_i^{(2)} = y_i^2. \tag{A6}$$

with the final linear transformation

$$\mathbf{w}^{(2)} = [1/4 \quad -1/4], \qquad b^{(2)} = 0. \tag{A7}$$

which gives

$$z_1 = 1/4\, y_1^2 - 1/4\, y_2^2 = x_1 x_2. \tag{A8}$$

Thus, the ASN implements multiplication exactly. Repeated composition of the same candidate functions can generate higher-degree polynomials, although the topology realizing a given expression is not unique.

# Appendix B: Effective-spin basis and construction of the benchmark Hamiltonian matrices

The Hamiltonian acts within the reduced Hilbert space spanned by direct products of the singlet and central-triplet states,

$$B^{ST} = \{||T_0\rangle^{(1)}, |S_0\rangle^{(1)}\} \otimes \{||T_0\rangle^{(2)}, |S_0\rangle^{(2)}\} \otimes \ldots \otimes \{||T_0\rangle^{(N)}, |S_0\rangle^{(N)}\} \quad \text{(B1)}$$

Within this reduced representation, the intrapair interaction appears as an effective Zeeman term proportional to $\hat{I}_{i,z}$, which lifts the degeneracy between the $|T_0\rangle^{(i)}$ and $|S_0\rangle^{(i)}$ states of the i-th $CH_2$ group. The identification $|\uparrow\rangle \equiv |T_0\rangle$ and $|\downarrow\rangle \equiv |S_0\rangle$ is used throughout the manuscript. The effective couplings $\Delta J_i$ describe nearest-neighbour interactions between adjacent effective spins. In the original aliphatic chain, these couplings are given by the difference between the two conformationally averaged vicinal three-bond *J*-couplings (see **Fig. B1**):

$$\Delta J_i \equiv \langle J_{i,i+3}\rangle - \langle J_{i,i+2}\rangle \quad \text{(B2)}$$

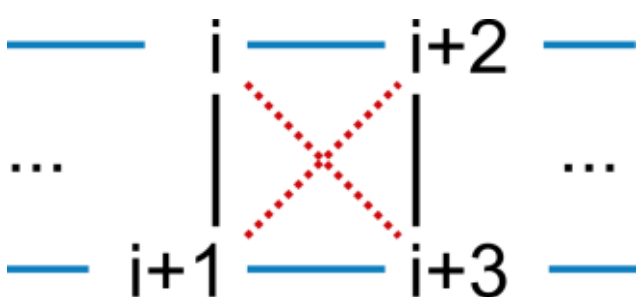

**Fig. B1. Proton-numbering convention for two neighboring methylene groups**. Protons $i$ and $i+1$ form the first pair, while $i+2$ and $i+3$ form the second. Here, $i$ labels individual protons rather than effective-spin sites.

The first term in Eq. (8) dominates under realistic conditions because the intrapair coupling $J_{intra}$ is substantially larger than all effective couplings $\Delta J_i$. Its eigenstates are direct-product states in the effective spin basis, such as $|\uparrow\uparrow\uparrow\downarrow\rangle$. The corresponding eigenenergies are obtained by assigning an energy contribution of $J_{intra}/4$ to each spin-up state and $-3J_{intra}/4$ to each spin-down state and summing these contributions over all sites:

$$E^{(0)} = \left(\frac{n_\uparrow}{4} - \frac{3n_\downarrow}{4}\right) J_{\text{intra}}, \qquad n_\uparrow + n_\downarrow = N. \quad \text{(B3)}$$

The exchange and double-quantum terms in Eq. (8) both generate matrix elements of $\Delta J_i/2$. Exchange terms connect product states differing by $|\cdots \uparrow\downarrow \cdots\rangle \leftrightarrow |\cdots \downarrow\uparrow \cdots\rangle$, thereby lifting their degeneracy and substantially modifying the eigenstates. This term is equivalent to the nearest-neighbour XY spin-chain Hamiltonian, whose eigenstates and eigenfrequencies are known analytically [9].

The double-quantum terms connect states differing by two spin flips: $|\cdots \uparrow\uparrow \cdots\rangle \leftrightarrow |\cdots \downarrow\downarrow \cdots\rangle$. Although the exchange and double-quantum matrix elements have the same magnitude, the exchange term acts within degenerate manifolds and contributes at first order. By contrast, the double-quantum term couples manifolds separated by an energy of order $2|J_{intra}|$ and therefore contributes to the eigenenergies only at second order when $|\Delta J_i|/2 \ll 2|J_{intra}|$. Its effect can consequently be treated perturbatively [8,9,12].

The exchange interaction preserves the number of $|\downarrow\rangle$ sites, whereas the double-quantum interaction changes it by ±2. Both interactions therefore preserve its parity, and the Hamiltonian separates into even- and odd-parity blocks, each of dimension $2^{N-1}$. For the three-site benchmark, **Fig. 2(a)** shows the four-dimensional even-parity block, comprising one state with no $|\downarrow\rangle$ sites and three states with two $|\downarrow\rangle$ sites. For the four-site benchmark, **Fig. 2(b)** shows the eight-dimensional odd-parity block, comprising four states with one $|\downarrow\rangle$ site and four states with three $|\downarrow\rangle$ sites.

# Appendix C: Architecture selection and construction of the extracted polynomial expressions

Candidate three-site architectures with different depths and widths were trained using the same activation library. Each candidate was characterized by its prediction error and by the number of arithmetic operations in the extracted symbolic expressions after coefficient thresholding. A candidate belonged to the Pareto front when no other model had both lower error and lower symbolic complexity. The architecture (2, 16, 4) was retained because it provided a compact point on this front while preserving the accuracy required for the subsequent symbolic analysis.

Although the selected model contains one hidden layer, it comprises two consecutive activation-selection stages. Consider a hidden variable formed after identity selection in the first stage,

$$h_s = w_{s1}r_1 + w_{s2}r_2 + b_s. \quad \text{(C1)}$$

If this variable selects the quadratic activation in the following stage, its contribution becomes

$$h_s^2 = w_{s1}^2 r_1^2 + 2w_{s1}w_{s2}r_1r_2 + w_{s2}^2 r_2^2 + 2b_s w_{s1} r_1 + 2b_s w_{s2} r_2 + b_s^2. \quad \text{(C2)}$$

A single quadratic activation therefore generates $r_1^2$, $r_2^2$, and the mixed term $r_1r_2$, together with linear and constant contributions from the bias. Other hidden variables selecting the identity function provide additional linear terms, while zero activations suppress unnecessary paths. Linear combinations of these hidden-node outputs produce the complete quadratic form recovered in Eqs. (12a)–(12d). If a first-stage quadratic output is squared again, the same architecture can generate quartic terms, although such terms were not retained in the extracted benchmark expressions.

# Appendix D: Two-dimensional degenerate perturbation theory for the nonuniform three-site chain

Within the even-parity block, let $P$ project onto the three-dimensional manifold spanned by $\{|\uparrow\downarrow\downarrow\rangle, |\downarrow\uparrow\downarrow\rangle, |\downarrow\downarrow\uparrow\rangle\}$, with unperturbed energy $A = -5J_{intra}/4$, and let $Q$ project onto the remaining state $|\uparrow\uparrow\uparrow\rangle$, with energy $B = 3J_{intra}/4$. In this ordered basis, the first-order perturbation is:

$$V_P = \begin{bmatrix} 0 & d_1 & 0 \\ d_1 & 0 & d_2 \\ 0 & d_2 & 0 \end{bmatrix}, \qquad d_i = \frac{\Delta J_i}{2}, \qquad R = \sqrt{d_1^2 + d_2^2}. \tag{D1}$$

The first-order eigenvalues are $+R$, 0, and $-R$, with normalized eigenvectors

$$|+\rangle = \frac{1}{\sqrt{2}R}\begin{bmatrix} d_1 \\ R \\ d_2 \end{bmatrix} |0\rangle = \frac{1}{R}\begin{bmatrix} d_2 \\ 0 \\ -d_1 \end{bmatrix}, \quad |-\rangle = \frac{1}{\sqrt{2}R}\begin{bmatrix} d_1 \\ -R \\ d_2 \end{bmatrix}. \tag{D2}$$

The Q-manifold state couples to P through the vector and squared matrix elements

$$\mathbf{v}_{QP} = [d_2 \quad 0 \quad d_1], \quad |\langle Q|V|\pm\rangle|^2 = \frac{2d_1^2 d_2^2}{R^2}, \quad |\langle Q|V|0\rangle|^2 = \frac{\left(d_2^2 - d_1^2\right)^2}{R^2}. \tag{D3}$$

Using $A - B = -2J_{intra}$ gives the second-order corrections within $P$,

$$\delta E_{\pm}^{(2)} = -\frac{d_1^2 d_2^2}{J_{intra} R^2}, \qquad \delta E_0^{(2)} = -\frac{\left(d_2^2 - d_1^2\right)^2}{2J_{intra} R^2}. \tag{D4}$$

whereas the Q-manifold state receives

$$\delta E_Q^{(2)} = \frac{R^2}{2J_{\text{intra}}}. \tag{D5}$$

Substituting $d_i = \Delta J_i/2$ and $S = \Delta J_1^2 + \Delta J_2^2$ yields Eqs. (10a)– (10d). At $S = 0$, the rational contributions are assigned their continuous zero-coupling limits. Along the positive uniform branch, $\Delta J_1 = \Delta J_2 = \Delta J \geq 0$, these expressions reduce to Eqs. (9a)–(9d) through second order. When the sign of $\Delta J$ is reversed, the two split branches exchange their order.

# Acknowledgments

KFS acknowledges support by l'Agence Nationale de la Recherche (ANR) on the project THROUGH-NMR (ANR-24-CE93-0011-01).

AI-use disclosure. During manuscript preparation, the authors used OpenAI ChatGPT (GPT-5, accessed in 2026) to assist with language revision, organization, and consistency checks involving notation, cross-references, and selected analytical expressions. The authors directed the queries, critically evaluated all suggestions, and verified the scientific statements and equations against analytical calculations, numerical diagonalization, and the underlying code. ChatGPT was not used to generate the research data or make final scientific judgments. The authors take full responsibility for the content of the manuscript.

# Data Availability Statement

Data Availability Statement. The code and datasets supporting the findings of this study are available in the NMRKan repository: A. Yu. Kharin, *NMRKan: NMR symbolic regression repository*, GitHub (2026), https://github.com/Alexankharin/NMRKan